\documentclass[aps,prl,showpacs,preprint]{revtex4-1}
\usepackage{epsfig}
\usepackage{graphics}
\usepackage{caption}
\usepackage{subcaption}
\usepackage{graphicx}
\usepackage{epstopdf}
\DeclareGraphicsExtensions{,.pdf,.png,.jpg,.gif}

\begin{document}

\title{Quantum turbulence by vortex stirring in a spinor Bose-Einstein condensate}

\author{B. Villase\~nor, R. Zamora-Zamora,  D. Bernal, and V. Romero-Roch\'in} 

\email{romero@fisica.unam.mx}

\affiliation{Instituto de F\'{\i}sica, Universidad
Nacional Aut\'onoma de M\'exico. \\ Apartado Postal 20-364, 01000 M\'exico D.
F. Mexico. }


\date{\today}

\begin{abstract}
We introduce a novel mechanism to develop a turbulent flow in a spinor Bose-Einstein condensate, consisting in the stirring of a single line vortex by means of an external magnetic field. We find that density and velocity fluctuations have white-noise power spectra at large frequencies and that Kolmogorov 5/3 law is obeyed in the turbulent region. As the stirring is turned off, the flow decays to an agitated non-equilibrium state that shows an energy bottleneck crossover at small length scales. We demonstrate our findings by numerically solving two-state spinor coupled 3D Gross-Pitaevskii equations. We suggest that this mechanism may be experimentally implemented in spinor ultracold gases confined by optical traps. 
\end{abstract}

\maketitle

Ever since Kolmogorov seminal ideas on classical turbulence \cite{Kolmogorov1}, stochastic and universal laws have been sought for and discovered to be obeyed by highly complex fluid flows. Among the most celebrated predictions is the so-called ``5/3" energy cascade. Although turbulence in general has remained as a fascinating topic of study both classically \cite{review-class} and in helium superfluids \cite{Vinen-review-old,Vinen-he}, there has been an enormous recent interest in the study of turbulence in ultracold superfluid gases, see Refs. \cite{Paoletti-review,Nemirovskii-review,Tsubota-review} for recent reviews. Of notorious relevance are the experimental realizations of quantum turbulence (QT) in a superfluid ultracold $^{87}$Rb gas in a pure magnetic trap \cite{Henn}, and in a 2D version confined by an annular trap \cite{Neely}. In the theoretical side, a lot of effort has been devoted to understand QT as a ``vortex tangle" as well as the velocity statistics of the turbulent state \cite{Kobayashi2,White1,White2,Seman,Nowak1}.

Following the findings of Ref. \cite{Us}, where it was shown that Gross-Pitaevskii (GP) spinor Bose-Einstein condensates (s-BEC) in inhomogenous optical traps can support vortices ``on demand" in the presence of external magnetic fields, we study here the dynamics generated by the simple stirring of a single vortex in a 3D, spin 1/2, s-BEC. 
We shall show that even for mild stirrings a turbulent flow is developed in both components of s-BEC. In what follows, we describe the system and the temporal excitation protocol, as well as the non-equilibrium stationary state that it is reached once the excitation is turned-off.  We discuss several scenarios.

As succinctly put in Ref. \cite{Allen}, the study of quantum turbulence should address the following three questions: i) the generation of turbulence; ii) the statistical steady state; and, iii) the decay of turbulence. In this Letter, we analyze a novel, simple and realizable stirring procedure to produce turbulence in ultracold gases.
The lack of a dissipative mechanism in GP superfluids makes it difficult to produce a true turbulent steady state, yet a turbulent phase is clearly obtained during and after the stirring process.  When the latter is turned off, one finds that the flow decays to a non-equilibrium stationary state with turbulence remnants and an energy bottleneck contribution at small length scales \cite{Lvov}. 
We point out the existence of other proposals for the creation of QT, either by starting in a non-equilibrium state \cite{Berloff} or by rotating a BEC cloud \cite{Kobayashi2,Parker}, all of them in a one-component BEC cloud. Ours necessarily requires a multicomponent spinor BEC. 

\begin{figure}
\centering
\includegraphics[width=0.5\textwidth]{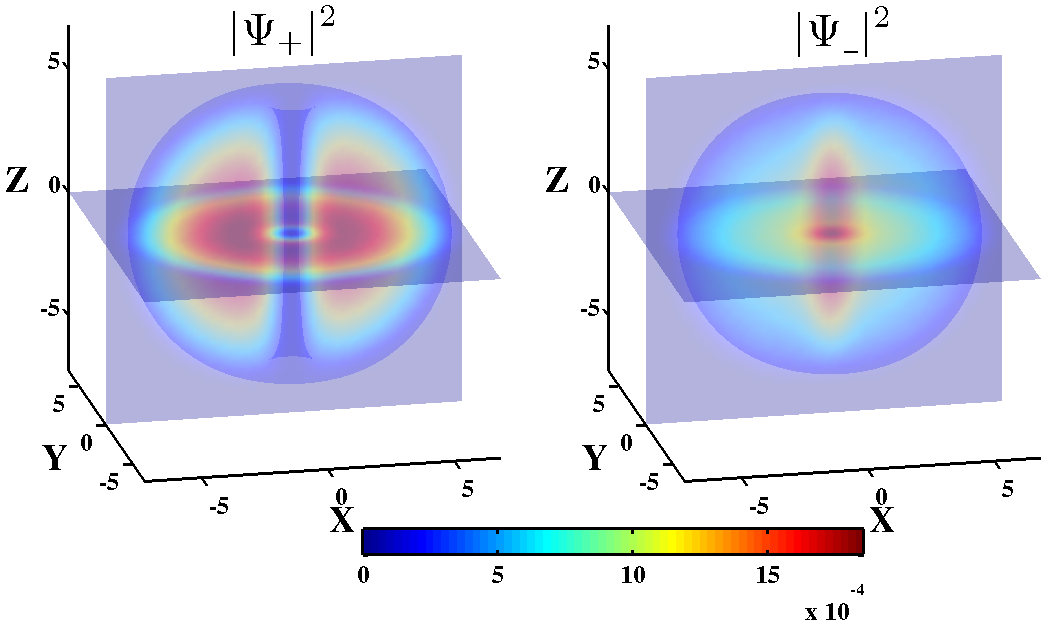}
\caption{(Color online) Density plots of the initial state. Component $\alpha = +$ shows a line vortex, of topological charge +1, along the $z-$direction at $x = y = 0$. Component $\alpha = -$  presents a density spike in the same spatial region where the vortex is located. }
\label{F1}
\end{figure}

We numerically solve the following pair of GP equations describing a 3D $s = 1/2$ s-BEC \cite{GPU,SM},
\begin{eqnarray}
[-\frac{\hbar^2}{2m} \nabla^2 + \frac{1}{2} m \omega^2 r^2 + g \left(\left|\psi_+\right|^2 + \left|\psi_-\right|^2\right) 
&&\nonumber 
\\  -m_0 B_x \sigma_x - m_0 B_y \sigma_y - i \hbar \frac{\partial}{\partial t} ]
\left(
\begin{array}{c}
\psi_+ \\ \psi_-
\end{array} \right) &=& 0 . \label{GPs}
\end{eqnarray}
The s-BEC is confined by an isotropic harmonic trap of frequency $\omega$, with atomic mass $m$, contact interactions independent of the spinor component $g = 4 \pi \hbar^2 a N/m$, with $a$ the common scattering length and $N$ the number of atoms. $\sigma_x$ and $\sigma_y$ are Pauli matrices. In dimensionless units $\hbar = m = \omega = 1$, we use $g = 8000$; this may represent $N = 10^5$ $^{87}$Rb atoms, with $a = 50$ \AA, confined in a dipolar optical trap with $\omega = (2 \pi) 100 $ Hz. These data are close to the actual values of the two hyperfine states $F = 1, m_F = -1$ and $F = 2, m_F = 2$ in $^{87}$Rb, see Ref. \cite{Egorov}. A dimensionless time unit corresponds to 1.6 ms approximately. In the above equations $m_0$ is the atom magnetic moment; we consider the magnetic field as,
\begin{equation}
m_0 B_x = \kappa \> x \>\>\>\>\>\>\> m_0 B_y = - \kappa \> \left( y - y_0 \sin \Omega t \right) \label{fields}
\end{equation}
where $\kappa = 1.0$ in dimensionless units, and $y_0$ and $\Omega$ are the amplitude and frequency of the temporal excitation. Initially, for times $t \le 0$, we set $y_0 = 0$ and let the system reach its stationary ground state. The solution is $\psi_\alpha (\vec r, t) = e^{-i \mu t/\hbar} \psi^0_\alpha(\vec r)$, where $\mu \approx 15.5$ is the chemical potential and $\psi^0_\alpha(\vec r)$ is a solution with a single vortex line of charge +1 in the $\alpha = +$ component at $x = y = 0$, and with a density spike in the same spatial region for component $\alpha = -$; see Ref. \cite{Us} for details of how this solution is obtained. One can also check that the expectation value of angular momentum is different from zero only in $z-$direction of the $\alpha = +$ component. Fig. \ref{F1} shows density plots of this vortex solution. 
We recall that, by writing $\phi_\alpha = |\psi_\alpha| e^{i \phi_\alpha}$, the density and velocity fields are given by $\rho_\alpha =  |\psi_\alpha|^2$ and $\vec v_\alpha = \hbar \nabla \phi_\alpha/m$. In Ref. \cite{Yukawa} a thorough discussion of the hydrodynamical formulation of spinor BEC is given.
 
At $t = 0$, the excitation is turned on, namely $y_0 \ne 0$ in Eq. (\ref{fields}). The excitation is quite simple, the line with zero magnetic field oscillates along the $y-$direction with amplitude $y_0$ and frequency $\Omega$. Nonetheless, the ensuing superfluid flow behavior is quite complex, depending on the values of $y_0$ and $\Omega$ and on the time the oscillation is kept on. We have made an extensive study of these variables (not shown here) and found that that there are threshold values of $y_0$ and $\Omega$, above which, a complicated turbulent flow pattern appears. We also find that angular momentum is excited in all spatial and spinor components. For the purposes of the present article, we have chosen the values $y_0 = 1.0$, about  1/7 the Thomas-Fermi radius of the condensate, and $\Omega$ equal to half the trap frequency $\omega$, this way avoiding resonances of collective motions \cite{Stringari}. 

We study three different excitation cases, (I) the excitation is never stopped; (II) the excitation is kept for a time $\tau_e = 5 (2 \pi)/\Omega$, then the external magnetic field is ramped down, $\kappa \to 0$, linearly in a time $0.4 \tau_e$; and (III) the excitation is kept for the time $\tau_e$, the field returns to its initial value but it is never turned off. Due to the lack of an energy dissipation mechanism, the excitation injects energy into the system and remains there. Therefore, in case (I) energy is kept entering the system, while in the others, energy is only received by the system during the excitation time. 
To be precise, the total energy we monitor is given by the following expression,
\begin{eqnarray}
 E &= &\int \left( -\frac{\hbar^2}{2m}\psi_{\alpha}^*\nabla^2\psi_{\alpha}+V_{ext}\psi_{\alpha}^*\psi_{\alpha} \right. \nonumber\\
 &&\left. - m_0 \psi_{\alpha}^* \left[ \vec{B} \cdot \vec{\sigma}\right ]^{\alpha\beta}\psi_{\beta}   + \frac{g}{2} \psi_{\beta}^{*}\psi_{\beta}\psi_{\alpha}^*\psi_{\alpha} \right) d^3 r. \label{energy}
 \end{eqnarray}
In cases (II) and (III) the energy is constant after the excitation stops and the system appears to reach a non-equilibrium stationary state. A way to determine such a stationary state is by monitoring the total average angular momentum (not shown here), which tends to a stable value after a somewhat long time of the order of $2 \tau_e$. As shown below, we analyze several statistical properties of the stationary state. 

\begin{figure*}
\centering
\includegraphics[width=0.8\textwidth]{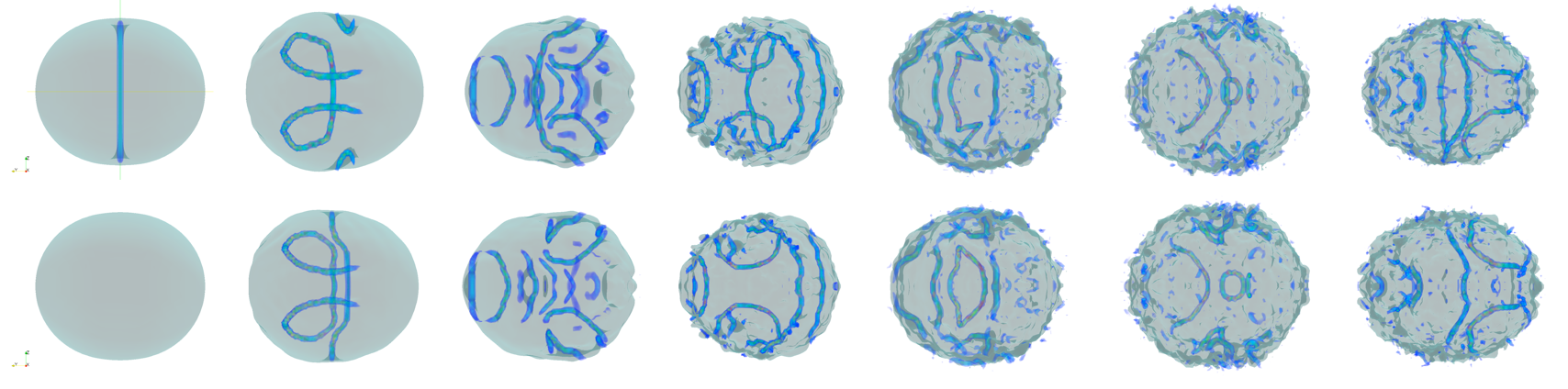}
\caption{Time evolution of the velocity fields of both spinorial components during the excitation time $0 \le t \le \tau_e$, with $\tau = 5 (2 \pi)/\Omega$. Upper row corresponds to component $\alpha = +$, lower one to $\alpha = -$. The snapshots are at times $t_n = n (\tau_e/6)$ with $n = 0, 1, \dots, 6$. The original line vortex bends and deforms, and another vortex appears at the other  component. In addition, pairs of line and ring vortices are nucleated in each component. See Supplemental Material \cite{SM} for full video.}
\label{F2}
\end{figure*}

\begin{figure*}
\centering
\begin{subfigure}[b]{0.4\textwidth}
\centering
\includegraphics[width=1.\textwidth]{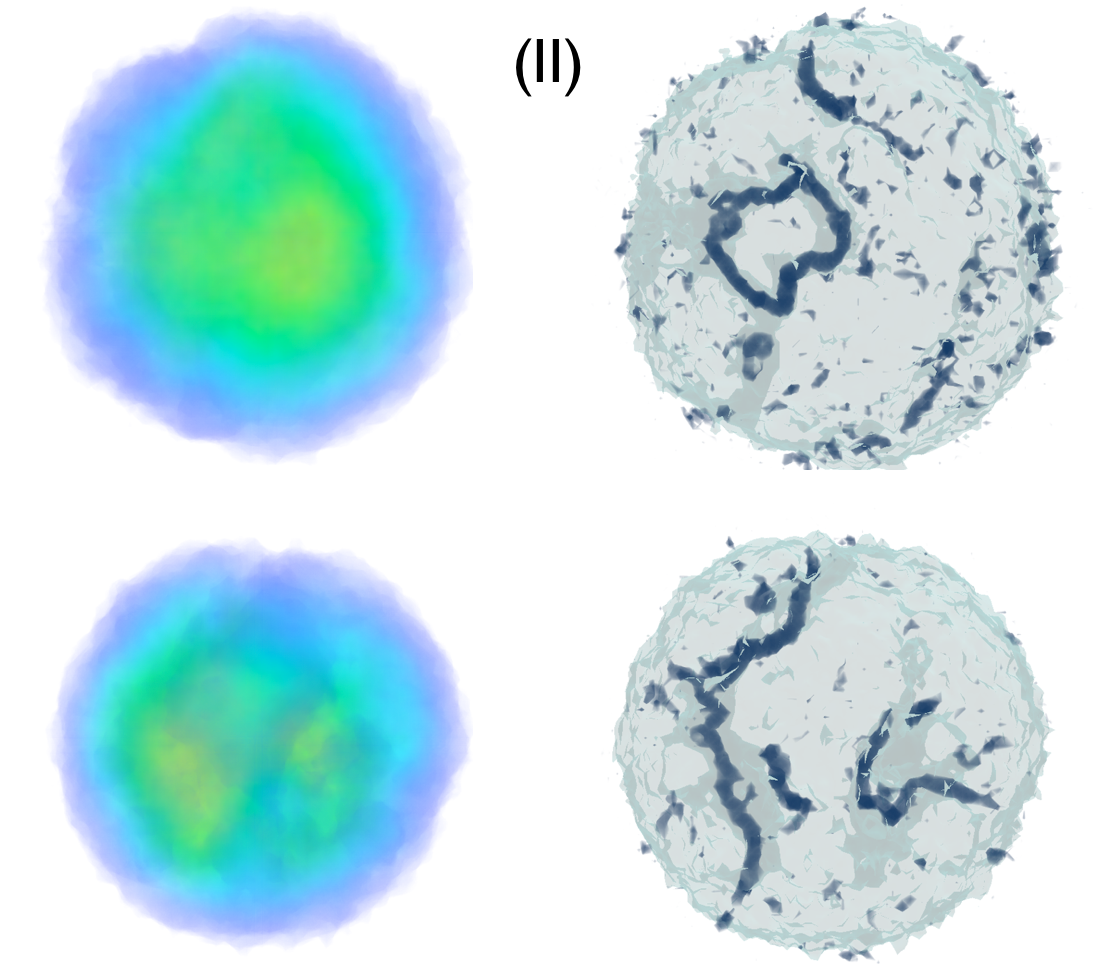}
\end{subfigure}
\begin{subfigure}[b]{0.4\textwidth}
\centering
\includegraphics[width=1.\textwidth]{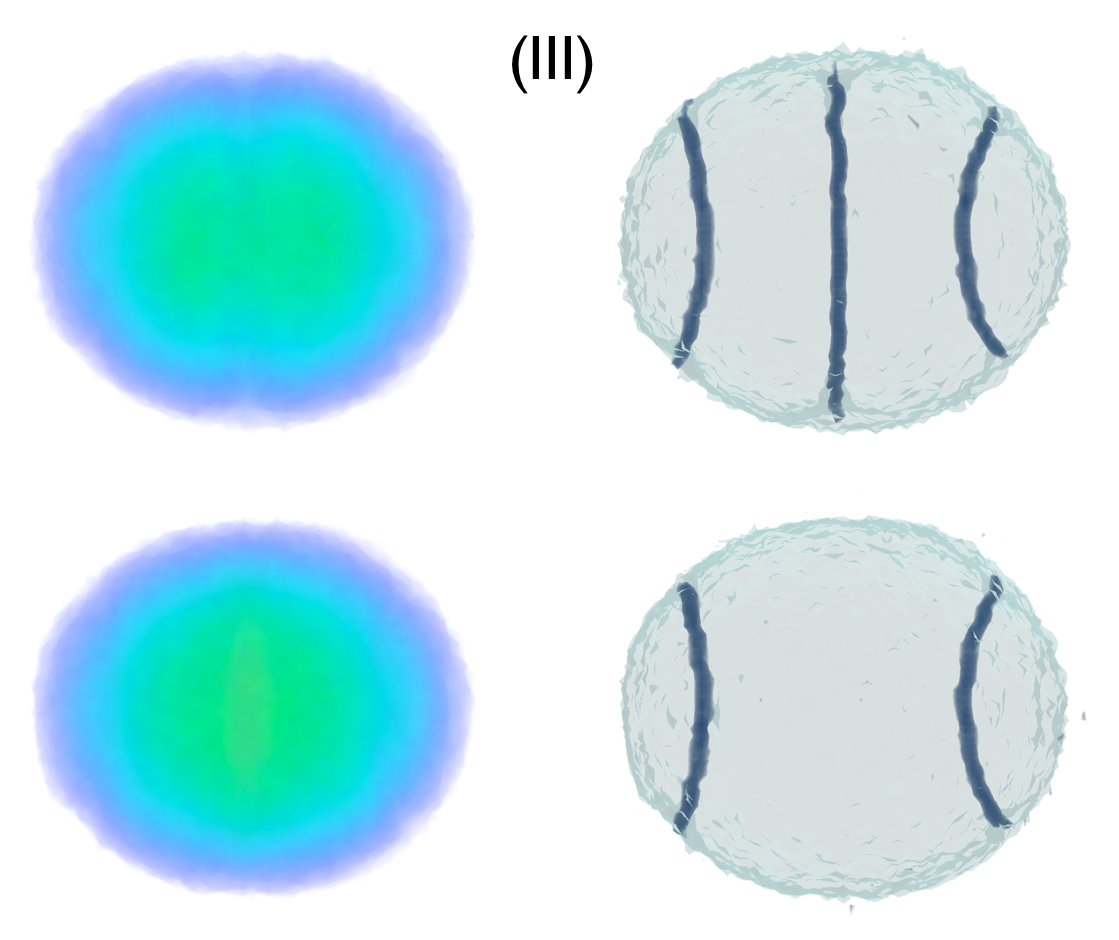}
\end{subfigure}
\caption{(Color online) Density and velocity fields after the stirring is turned-off; upper row is spinorial component $\alpha = +$, lower one for $\alpha = -$. In case (II) the external magnetic field is ramped out; the snapshot is at $t = 245$ units of time. In case (III) the magnetic field is kept on; the  snapshot is at $t = 366$ units of time.  In case (II) there appear two additional vortices, in each component, all with negative topological charge, that orbit at different frequencies around the central vortex. See Supplemental Material for full videos \cite{SM}.}
\label{F3}
\end{figure*}

In Figs. \ref{F2} and \ref{F3}, we show a series of snapshots of the time evolution of s-BEC under the protocols described above. In Fig. \ref{F2} we show the evolution during the excitation time $0 \le t \le \tau_e$. We show velocity plots as seen from the $x-$direction. Full videos of this figures are shown in the Supplemental Material \cite{SM}. One can see that the superfluid flow cannot follow the motion of the magnetic field, creating a complicated pattern of line and ring vortices. As one can observe in the panels of Fig. \ref{F2}, taken at different times, first, the line vortex at component $+$ bends while another one is nucleated in component $-$. At the same time, vortices are created in pairs, one in component $+$ and another very similar one in the other component.  There are also two types of vortices, vortex rings and line vortices that end at the surface of the cloud. Because vortices nucleate, move and seem to disperse and dissipate at the surface of the cloud, it is very difficult to quantify the total circulation. Since the fluid is compressible, Kelvin waves and phonon excitations appear to  be generated creating bursts in the velocity field. Although a naive idea about a inviscid flow might lead us to believe that a portion of a superfluid cannot efficiently drag its surroundings, one clearly sees that this is not the case; namely, the fluid appears as being effectively stirred by the motion of the magnetic field, and its lagging behind creates a wake that breaks into the observed vortex and phonon excitations. This, unavoidably, creates a turbulent state. The same mechanism appears to be responsible for the decay to an non-equilibrium steady state once the stirring is turned-off.

In Fig. \ref{F3} we show snapshots of the flow of cases (II) and (III), at a time later than the excitation time $t \gg \tau_e$; full videos of the whole evolution are also shown in the Supplemental Material \cite{SM}.  In case (II), while there is no magnetic field in the steady state, the  flow appears to remain agitated due to the lack of a dissipation mechanism. Case (III) is particularly interesting because it shows a stationary semi-turbulent state with two additional line vortices in each spinorial component, with topological charge -1 all of them, in addition to the original vortex at the origin, keeping its  initial charge +1. The additional vortices orbit around the original vortex with different angular velocities. In case (I), where the excitation is never stopped, the most notorious aspect is that, while vortices and phonons keep being excited, the velocity fields acquire maximum values at the edge of the clouds.

\begin{figure*}
\centering
\begin{subfigure}[b]{0.4\textwidth}
\centering
\includegraphics[width=1.\textwidth]{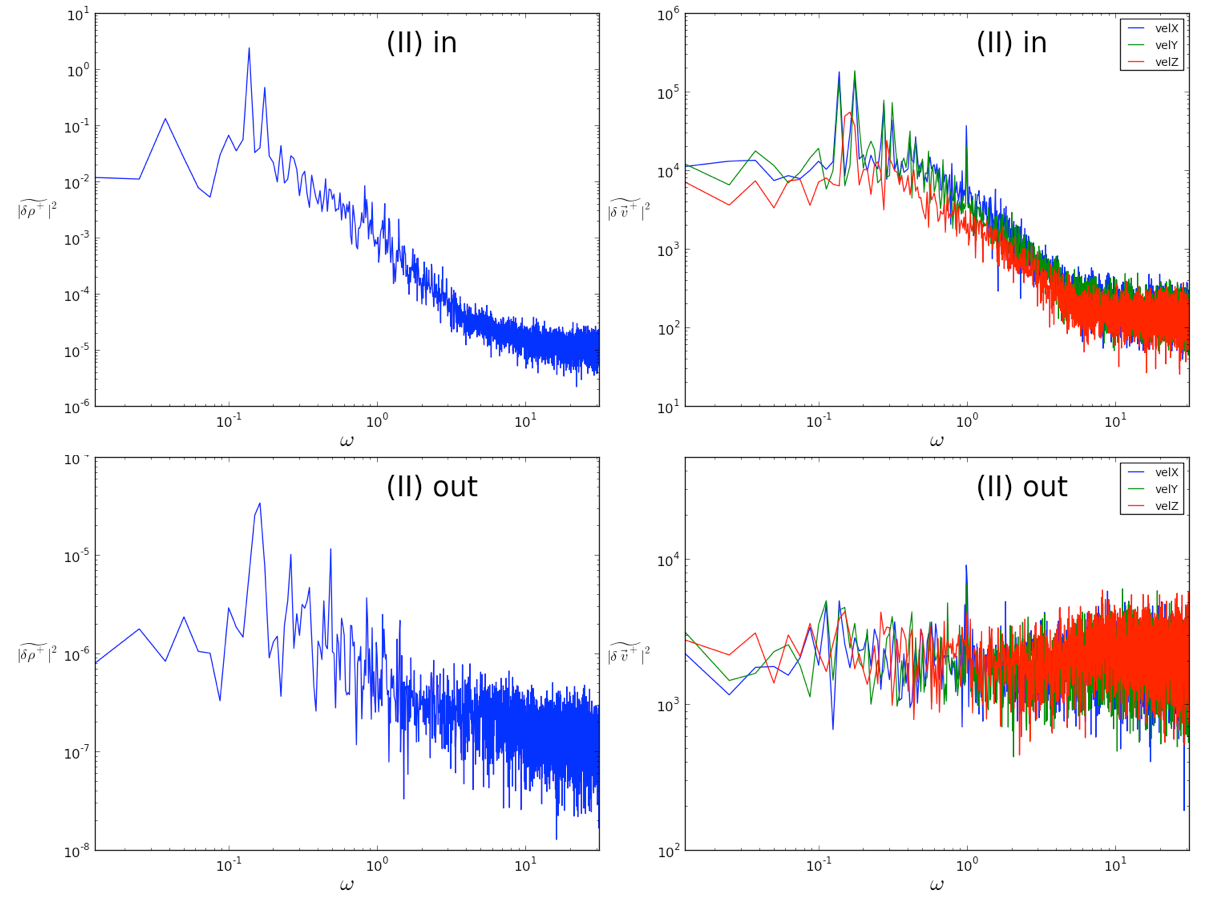}
\end{subfigure}
\begin{subfigure}[b]{0.4\textwidth}
\centering
\includegraphics[width=1.\textwidth]{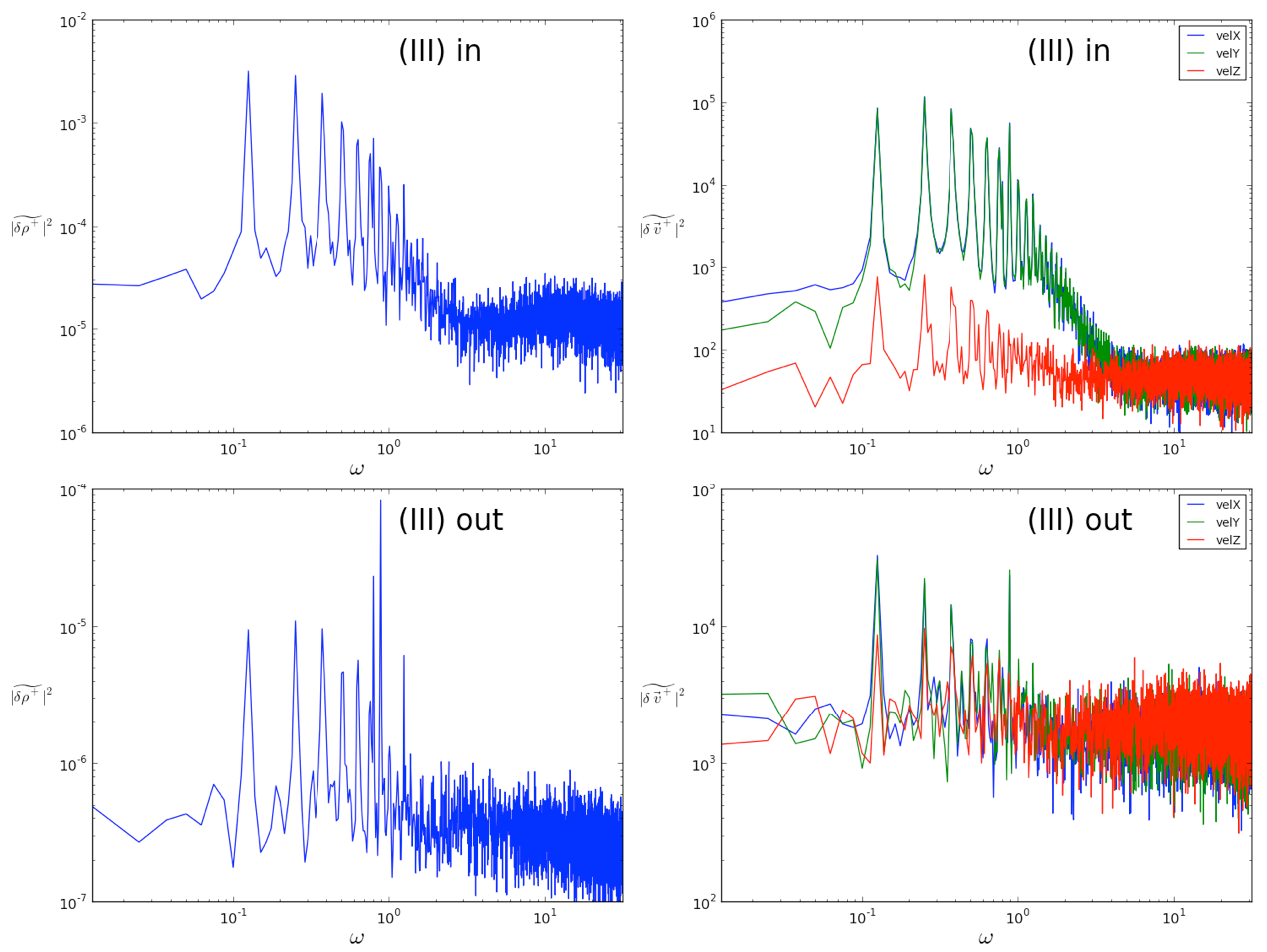}
\end{subfigure}
\caption{(Color online) Power spectra of density $|\widetilde {\delta  \rho}_\alpha(\vec r, \omega)|^2$ and velocity $|\widetilde {\delta  {\vec v}}_\alpha (\vec r, \omega)|^2$, for $\alpha = +$, of the non-equilibrium steady state, for two spatial positions for cases (II)  and (III). The label ``in" denotes a position near the original vortex line, while ``out" to a position near the edge of the cloud. }
\label{F4}
\end{figure*}

To describe the statistical properties of the stationary state in cases (II) and (III), we consider the following strategy. First, after $t = \tau_e$, we let the system evolve until the magnitude of the total angular momentum decays to an almost constant value, signaling a ``stationary" state; this time is of the order of $2.0 \tau_e$. Then, we let the system evolve another long interval of time, of the order of $6.5 \tau_e$, during which we calculate the time average of the density and velocity fields, as well as their fluctuations; that is, 
\begin{eqnarray}
\rho_\alpha(\vec r,t) &=& \rho_\alpha^s(\vec r) + \delta \rho_\alpha(\vec r,t) \nonumber \\
\vec v_\alpha(\vec r,t) &=& \vec v_\alpha^s(\vec r) + \delta \vec v_\alpha(\vec r,t)
\label{fluc}
\end{eqnarray}
where the ``$s$" superscript denotes the time-average of the quantity in question. The idea is that, barring low-frequency collective modes, the fluctuating part should capture the expected universal and homogenous turbulent contribution. In Fig. \ref{F4}, we present the power spectrum of the fluctuating part, both for density $|\widetilde {\delta  \rho}_\alpha(\vec r, \omega)|^2$ and velocity $|\widetilde {\delta  {\vec v}}_\alpha (\vec r, \omega)|^2$ fields, for the + spin component of (II) and (III) cases, and for two different spatial regions; one in the vicinity of the position of the original line vortex, and the other near the edge of the cloud. The most relevant result is that, the farther the spatial point is from the initial vortex region, the power spectra is essentially white noise. Near the vortex region the spectra shows both white-noise in the ultraviolet region while some kind of $1/f^\beta$ noise in the low frequency region, with $\beta \approx 0-2.5$. This latter behavior may be ``contaminated" from low frequency collective modes that cannot be completely removed with the time average. In case (III) the large peaks at low frequency correspond to the orbiting motion of the stable vortices, see Fig. \ref{F2}. The white-noise fluctuating contribution may be identified as the remnants of the turbulent component of the flow.

\begin{figure*}
\centering
\begin{subfigure}[b]{0.3\textwidth}
\centering
\includegraphics[width=1.\textwidth]{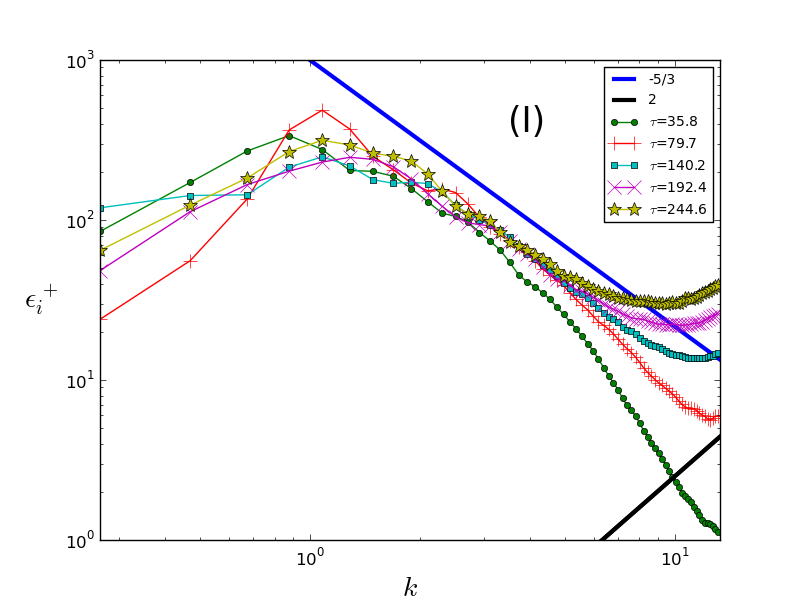}
\end{subfigure}
\begin{subfigure}[b]{0.3\textwidth}
\centering
\includegraphics[width=1.\textwidth]{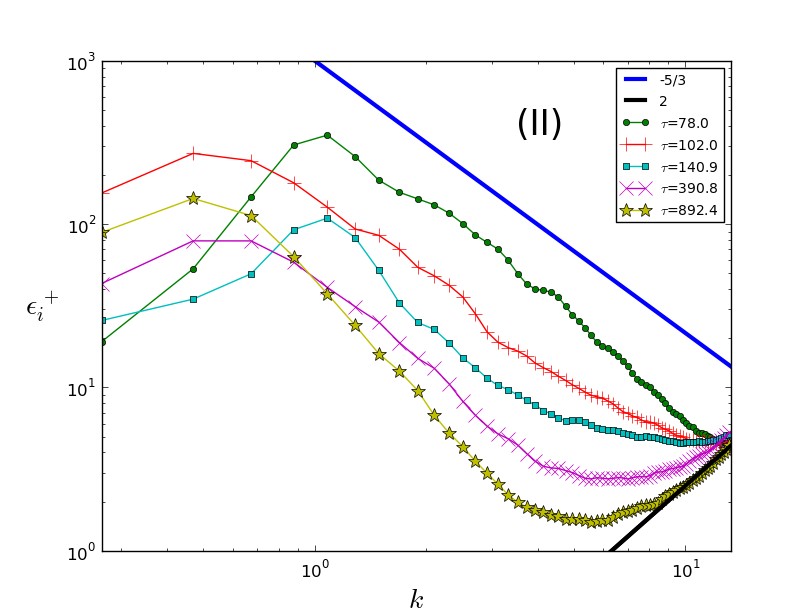}
\end{subfigure}
\begin{subfigure}[b]{0.3\textwidth}
\centering
\includegraphics[width=1.\textwidth]{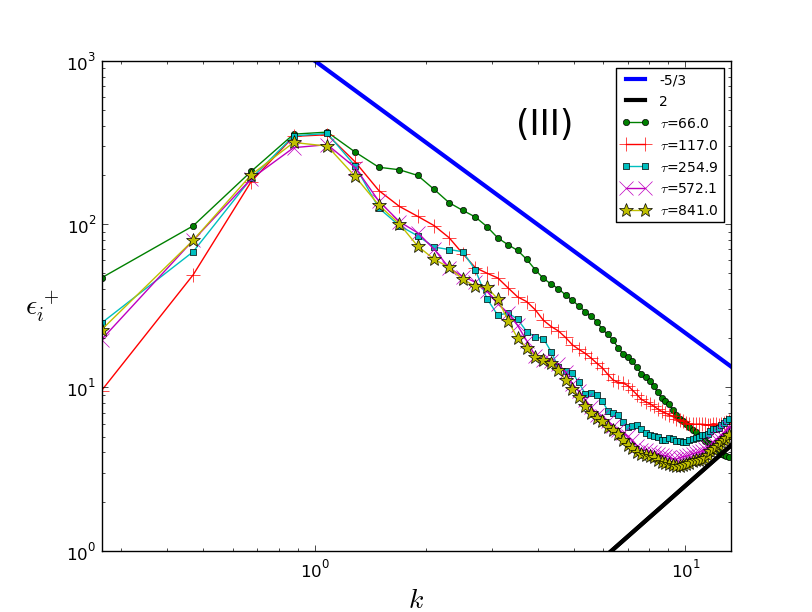}
\end{subfigure}
\caption{(Color online) Incompressible kinetic energy $\epsilon_i^{+}$ vs. $k$ for cases (I) - (III), taken at different times. The (blue) solid line is at slope $-5/3$, while the (black) solid one is at slope $+2$.}
\label{F5}
\end{figure*}

The second analysis concerns the wavenumber dependence of the incompressible part of the kinetic energy. It is widely believed \cite{Vinen-he,Paoletti-review,Nemirovskii-review,Tsubota-review} that in the turbulent region such a dependence should show Kolmogorov ``5/3" law \cite{Kolmogorov1}, indicating an energy cascade from large to small length scales. This is also true in wave turbulence \cite{Nazarenko}. Since GP superfluid flow has both compressible and incompressible contributions, we decompose the kinetic energy into both parts \cite{Nore} and calculate the wave vector dependence of the component $\alpha$, defined as,
\begin{equation}
K_i^\alpha = \int \> \frac{1}{2} \left( \rho_\alpha \vec v_\alpha^2 \right)_i \> d^3r = \int \> \epsilon_i^\alpha(k) \> dk \label{K}
\end{equation}
where $i$ stands for incompressible and $k$ is the magnitude of wavector $\vec k$. The expected Kolmogorov law is $\epsilon_i^\alpha(k)  \sim k^{-5/3}$ in an intermediate range of $k$, corresponding to length scales $l < \lambda < a$, where $l$ and $a$ are of the order of the sizes of the cloud and of the vortices core. In Fig. \ref{F5} we show $\epsilon_i^{+}(k)$ for the three cases (I)-(III) and for several times, as indicated in the figure. We have several observations. To begin with, one sees that the shape of the curve is reminiscent of the typical ones of classical turbulence \cite{review-class} for small $k \lesssim l^{-1}$, then it has a region that obeys Kolmogorov law (at least for not too long times), but then it differs from the classical one in the tail $k \sim a^{-1}$. In the classical ones, the curve bends down indicating the region where viscosity plays its main role dissipating the energy into heat. Here, we see that the curve bends up for long $k$, as $\epsilon(k) \sim k^2$, a free particle-like spectrum. We believe this has an interesting explanation. First of all, it is clear that an energy cascade is present: energy enters the system through the stirring of the initial line vortex, creating in turn smaller vortices and other compressible excitations as phonons. This transfers the energy to smaller length scales. However, since GP does not have a mechanism to dissipate heat into a thermal cloud, energy tends to be concentrated at high-$k$ excitations. This concentration may lead to a state of quasi-equilibrium with a kind of equipartition of energy yielding the $k^2$ dependence; a similar contribution has been predicted to appear in wave-turbulence of Kelvin waves, as a bottleneck phenomenon \cite{Lvov}. At the same time, however, we recall that as time passes by, most of the large fluctuations occur at the edges of the clouds (as can be seen in the videos of the Supplemental Material \cite{SM}):  It is as if the excitations not only cascade their energy but also migrate it to the outer regions of the cloud. This has a further consequence. In real experimental finite-temperature BEC clouds, the superfluid region is concentrated in the center of the cloud, while the outer region is a thermal shell where viscosity and heat dissipation can occur. This part is completely missed by GP. However, our calculations indicate that, if the thermal cloud were included, the small length scale excitations would find their way there and would eventually be dissipated into heat. This would be a turbulence decay mechanism, with a concomitant increase of temperature of the whole BEC cloud. A calculation of this sort may be implemented with the techniques of Refs. \cite{Kobayashi2,Kobayashi,Jackson,Gardiner}. Regarding Kolmogorov law, we find that it is better obeyed in intermediate times, when turbulence is better observed, and that for longer times the long-$k$ accumulation tends to bend the curve, deviating it slightly from the -5/3 slope. Because our simulated BEC clouds are not very large \cite{GPU}, it is not clear whether this is deviation is a true or a finite-size effect.

To conclude this Letter, we insist that the present study indicates a very feasible, simple and novel way of producing quantum turbulence in ultracold  superfluids. First, in many laboratories, it is now almost routine to produce spinorial BEC in dipolar optical traps, and second, as described in Ref. \cite{Us}, the initial line vortex may be fairly simple produced by an arrangement of long wires carrying appropriate electric currents. The stirring may be performed by using AC currents.\\

\newpage

\centerline{\bf \Large Supplemental Material}

\medskip
{\bf Numerical procedure}

The numerical solutions of the two GP coupled equations, Eqs. (1) in the main text, were performed using a 4th order Runge-Kutta method for the time evolution and a finite-difference procedure for the evaluation of the laplacian terms. The initial state was obtained with the relaxation method of Ref. \cite{Zeng}. The code was implemented to be performed in parallel in graphic processor units (GPU); in particular, the finite-difference method was implemented to exploit the GPU's textures memory. The code was developed in {\it py-CUDA} \cite{pyCUDA}, a variation of {\it Python} \cite{Python} that includes {\it CUDA} \cite{CUDA}. The calculations were performed in a Tesla C2075 GPU with 532 cores \cite{Tesla}. A typical run of $3.5 \times 10^5$ time steps takes nearly four hours computer time. We estimate that our calculation is approximately 80-fold faster than in a serial implementation of the program in a single CPU. The equations were solved in a $128^3$ spatial grid. Using dimensionless units with $\hbar = m = \omega = 1$, the length of a side of the calculation box is $L = 30$,  and the time step $\Delta t = 0.002$. For comparison, the Thomas-Fermi radius of the cloud is $R_F \approx 7$.

\medskip

{\bf Time evolution of cases (I), (II) and (III)}

As indicated in the main text, we present here full videos of the time evolution of the three excitation cases, (I) the excitation is never stopped; (II) the excitation is kept for a time $\tau_e = 5 (2 \pi)/\Omega$, then the external magnetic field is ramped down, $\kappa \to 0$, linearly in a time $0.4 \tau_e$; and (III) the excitation is kept for the time $\tau_e$, the field returns to its initial value but it is never turned off. The excitation time is $\tau_e \approx 62.8$ units of time. The videos are labeled {\bf Video I}, {\bf Video II} and {\bf Video III}, with their respective URL addresses given below. In each video we show 8 frames, the upper 4-ones corresponding to component $\alpha = +$, the lower 4-ones to $\alpha = -$. In each row we show one density plot, as seen from the $z$-axis, and three velocity plots, as seen from the three axis. The velocity plots show the regions where the magnitude of the velocity is larger than a given threshold, thus, the vortices are enhanced. In Fig. \ref{color} below, we show a color-coded bar for the density plots. On top of each video there is bar indicating the elapsed time. The ``Excitation" label refers to the situation when the external field is oscillating; the ``Relaxation" is an estimated time during which the expectation value of the angular momentum relaxes to an approximately constant value; and the ``Study time" to the interval during which fluctuations power spectra of  Fig. 4 were  calculated. Below we give a brief description of what may be observed in each video. 

\begin{figure}
\centering
\includegraphics[width=0.5\textwidth]{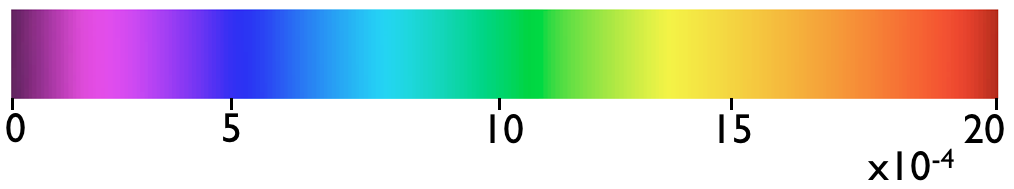}
\caption{Color-coded bar for the density plots in all videos below, in dimensionless units of number density. }
\label{color}
\end{figure}

\medskip
First, we note that during the excitation time $\tau_e 5 (2 \pi/\Omega) = 62.8$ units of time, all the videos look the same. During this time, the external field oscillates at frequency $\Omega = \omega/2$ with an amplitude $y_0 = 1.0$, see Eq. (2) of the main text. As the superfluid flow cannot follow the motion of the magnetic field, it creates a complicated pattern of line and ring vortices. It appears as if vortices are created in pairs, one in component $+$ and another very similar one in the other component.  There are also two types of vortices, vortex rings and line vortices that end at the surface of the cloud. One can also see that excitations of all types tend to accumulate at the surface of the cloud. The patter is obviously symmetric with respect to the plane $z = 0$.

\medskip
{\bf Video I}

http://www.youtube.com/watch?v=Ggp-Oc6e9lQ

\medskip
Because in this case the excitation never stops, energy keeps entering the system and the flow never reaches any kind of stationary state, due to its lack of a dissipative mechanism. One sees excitations emerging continuously and vortices seem to appear and disappear at the edges of the cloud. At the same time, one sees localized excitations, where the velocity is large, mostly accumulating at the edges of the cloud. We believe these are acoustic excitations. At the final stages of the video the images hardly allows to discern if there are vortex tangles inside the cloud, since most of the velocity excitations, as mentioned, are accumulated at the borders of the clouds.

\medskip
{\bf Video II}

http://www.youtube.com/watch?v=UftnkLtRWbU

\medskip
In this case, the magnetic field is turned-off, namely, $\kappa$ is ramped down linearly to $0$. This is indicated in the yellow section of the time-bar marked as ``Off". We can see that as long as $\kappa \neq 0$ the two components show the same vortices plus the original vortex that is only present in one and that keeps switching between components. When $\kappa$ reaches $0$ at the beginning of the green section of the time bar marked as ``Relaxation" both components show a similar behavior, but after a while they start showing an independent evolution. This is clearer in the ``Study Time" (in red on the time bar) where one can see that when a ring vortex is formed in one component there is no trace of it in the other, rather it may present linear vortices as seen in $t \approx  250$. Another aspect worth mentioning is that it seems that the energy  is accumulated at the edge of the cloud, by the end of the ``Excitation Time". This is seen as purple spots all around the vortices that tend to fade towards the end of the video, where we can see simply sporadical formation of vortices without the purple spots. This is a non-equilibrium stationary state.

\medskip
{\bf Video III}

http://www.youtube.com/watch?v=3cr5TQNVoeM

\medskip

In this case the magnetic field oscillates for $t_e = 62.8$ units of time, then it returns to its initial state and remains unchanged for the rest of the calculations. One can observe that, although the fluid appears agitated, a clear stationary state is reached in which in component spin $\alpha = +$, a vortex of charge +1 stabilizes at that the location of the original vortex; in addition two other line vortices along the $z$-direction, of charge -1 each, orbit around the central one, with different angular velocities. One can also see that two other very similar vortices are nucleated in spin component $\alpha = -$, also of charge -1 each one, and orbiting with the same angular velocities.

We thank support from grant DGAPA UNAM IN108812.

\end{document}